\documentstyle[12pt,axodraw]{article}
\begin{document}
\newcommand{\bq}{\begin{equation}}
\newcommand{\eq}{\end{equation}}
\newcommand{\bqa}{\begin{eqnarray}}
\newcommand{\eqa}{\end{eqnarray}}
\newcommand{\nl}{\nonumber \\}
\newcommand{\noi}{\noindent}
\newcommand{\epl}{e^+}
\newcommand{\emn}{e^-}
\newcommand{\into}{\;\;\to\;\;}
\newcommand{\order}[1]{{\cal O}\left(\alpha^{#1}\right)}
\newcommand{\ubar}{\bar{u}}
\newcommand{\ww}[2]{\langle #1 #2\rangle}
\newcommand{\wws}[2]{\langle #1 #2\rangle^{\star}}
\newcommand{\bfig}{\begin{center}\begin{picture}}
\newcommand{\efig}[1]{\end{picture}\\{\small #1}\end{center}}
\newcommand{\flin}[2]{\ArrowLine(#1)(#2)}
\newcommand{\wlin}[2]{\DashLine(#1)(#2){2}}
\newcommand{\zlin}[2]{\DashLine(#1)(#2){5}}
\newcommand{\glin}[3]{\Photon(#1)(#2){2}{#3}}
\newcommand{\lin}[2]{\Line(#1)(#2)}
\newcommand{\sof}{\SetOffset}
\newcommand{\bmip}[2]{\begin{minipage}[t]{#1pt}\bfig(#1,#2)}
\newcommand{\emip}[1]{\efig{#1}\end{minipage}}
\newcommand{\putk}[2]{\Text(#1)[r]{$p_{#2}$}}
\newcommand{\putp}[2]{\Text(#1)[l]{$p_{#2}$}}

\pagestyle{empty}
\begin{flushright}
INLO-PUB-12/94\\
July 1994
\end{flushright}
\vspace{2cm}
\begin{center}\begin{Large}\begin{bf}
EXCALIBUR -- a Monte Carlo program \\
to evaluate all four fermion processes \\
at LEP 200 and beyond \footnote{This research has been partly supported
by EU under contract number CHRX-CT-92-0004.}
\end{bf}\end{Large}\\
\vspace{\baselineskip}
{\bf  F.A. Berends and R. Pittau}\\
\begin{it} Instituut-Lorentz, University of Leiden,\\
P.O.B. 9506, 2300 RA Leiden, The Netherlands
\end{it}\\
\vspace{\baselineskip}
{\bf R.~Kleiss}\\
\begin{it} NIKHEF-H,\\
P.O.B. 41882, 1009 DB Amsterdam, The Netherlands \end{it}\\
\vspace{\baselineskip}
{\bf Abstract}
\end{center}
A Monte Carlo program is presented that computes all four fermion
processes in $e^+ e^-$ annihilation. QED initial
state corrections and QCD contributions are included.
Fermions are taken to be massless, allowing a very fast evaluation of
the matrix element. A systematic, modular and self-optimizing
strategy has been adopted for the Monte Carlo integration, that
serves also as an example for further event generators
in high energy particle physics.
\begin{center}
{\it submitted to Computer Physics Communications}
\end{center}

\newpage
\pagestyle{plain}
\setcounter{page}{1}
\begin{center}
{\Large {\bf Program Summary}}
\end{center}

\noi{\it Title of program:\/} EXCALIBUR\\

\noi{\it Program obtainable from:\/} R.~Kleiss, NIKHEF-H,\\
 P.O.B. 41882, 1009 DB Amsterdam, The Netherlands,\\
 t30@nikhefh.nikhef.nl\\
 R.~Pittau, Instituut-Lorentz, University of Leiden,\\
 P.O.B. 9506, 2300 RA Leiden, The Netherlands,\\
 rulgm0@leidenuniv.nl\\

\noi{\it Programming language used:\/} {\tt FORTRAN 77}\\

\noi{\it Memory required:\/} about 170kbytes\\

\noi{\it number of bits per word:\/} 32\\

\noi{\it Subprograms used:\/} none\\

\noi{\it Number of lines in distributed program:\/} 3784\\

\noi{\it Keywords:\/}
Decaying vector-boson production, all four fermion processes, electroweak
and QCD background, initial state QED radiation, multichannel Monte Carlo
approach.\\

\noi{\it Nature of physical problem:\/}
Heavy vector boson production will be investigated at $e^+ e^-$ colliders
in a wide range of energies. At LEP II, the relevant process
is
\bq
e^+e^-  \;\;\to\;\; W^+\;W^- \:\: .\label{wpair}
\eq
At higher energies other processes like
\bqa
e^+e^-  &\;\;\to\;\; & Z\;Z\;\;, \label{zpair}               \\
e^+e^-  &\;\;\to\;\; & W\;e\;\nu_e\;\;, \label{singlew}      \\
e^+e^-  &\;\;\to\;\; & Z\;e^+e^-\;\;, \label{singleze}     \\
e^+e^-  &\;\;\to\;\; & Z\;\nu_e\;\bar{\nu}_e\;\;, \label{singleznu}
\eqa
become important.
The detected experimental signal for all above processes
is a four fermion final state. Therefore, a Monte Carlo program being
able to take into account both
signal and background electroweak diagrams for {\em all} four fermion
processes is required. QED initial state radiation and QCD background
play also an important r\^ole and have to be included.\\

\noi{\it Method of solution:\/}
An {\em event generator} is the most suitable choice for a program to be able
to deal with the above physical problem, since each generated event is a
 complete description of the momenta of the produced particles and any
experimental cut can be easily implemented.
There are two basic difficulties. First of all the number of
Feynman diagrams can be very large. Secondly, taking into account also
the background diagrams, the peaking structure of the matrix element squared
is very rich, so that a straightforward integration over the allowed phase
space is impractical. The former problem can be solved by using spinorial
techniques to compute the amplitudes and taking massless fermions.
The latter requires the use of a {\em multichannel} approach, where the
integration variables are generated according to distributions that
approximately reproduce the peaking behaviour of the integrand, so reducing
the estimated Monte Carlo error.

Since one wishes to take into account {\em all} possible final state
(that means to have from 3 to 144 different Feynman diagrams, many
of them leading to different peaks in the phase space), a systematic
and automatic procedure for both the generation of the Feynman diagrams
and the phase space integration is unavoidable, together with an
algorithm for the self-optimization of the predetermined probabilities
used to choose the various {\em channels}.

All that has been implemented in {\tt EXCALIBUR}.
This paper serves also as an example of the entire procedure to be used
to build future {\em event generators}.\\

\noi{\it Typical running time:\/} about 100 events per second on HP,
     depending on the chosen physical process.\\

\noi{\it Unusual features of the program:\/} none

\vspace*{2cm}
\clearpage
\begin{center}
{\Large {\bf Long Write-Up}}
\end{center}
\section{Introduction}
In the near future LEP II will become operative in
the energy region around 200 GeV. The physics relevant at higher energies
will be investigated at the next generation of $\epl \emn$ linear
colliders.
Many interesting physics issues can be studied and
one of them is the gauge-boson production.
Around 200 GeV events with the signature of two produced $W$'s have a
large
cross section, while single boson production processes become important
with increasing energy \cite{hag}.
One can distinguish five sizeable reactions
(eqs. (\ref{wpair})-(\ref{singleznu}))
in which gauge bosons are produced.
Due to the fact that the massive bosons are unstable particles, all
those processes end up with a detectable 4-fermion final state to which
many Feynman diagrams can contribute. Some of them are related to the
reactions (\ref{wpair})-(\ref{singleznu}) (signal diagrams);
others are not (background diagrams).
For this reason a precise knowledge of all possible processes
\bq
\epl\;\emn \into \mbox{4 fermions}\;\; \label{fourferm}
\eq
is unavoidable in order to make comparison with experiment \cite{np}.

In addition to these background effects, one wants to be able to study
{\em any} experimental distribution, taking into account the
dominant radiative corrections effects, and the possibility to
implement {\em any} experimental cut.
To solve these problems we wrote
an event generator, that can handle all diagrams leading to
a specified 4-fermion final state (with, of course, the option of
a restriction to the signal diagrams), and that incorporates the LL
$\order{}$ and $\order{2}$ initial state radiation (ISR), with exponentiation
of
the remaining soft-photon effects \cite{qed}.
Furthermore, with a four quark final state, QCD diagrams are present
as well, giving non-negligible effects that have been also
included \cite{qcd}.

It should be noted that even when cross sections do not dramatically change
under
inclusion of tiny effects, there are quantities that are
very sensitive to any small correction. Among them is
the average energy $\epsilon$ radiated by the beams.
A precise knowledge of $\epsilon$ is required  at LEP II when the
reconstruction of the jet
invariant mass distributions is performed to measure the
$W$ mass \cite{qcd,katsa}.
In addition, $\epsilon$ is also very sensitive to the imposed
experimental cuts \cite{qed}, so that, once more, a Monte Carlo
approach is to be preferred.

In order to build a fast program we have taken the limit of vanishing fermion
masses.
Even if this implies the {\em absence} of diagrams where a Higgs boson
couples to the fermions - and therefore we cannot compute the Higgs signal -
we can
at least estimate the background. On the other hand,
the inclusion of the leading Higgs signal is trivial because only few
diagrams account for it and,
due to their helicity structure, do not interfere
with all the others in the limit of massless fermions.
However, this has not been implemented.

In the rest of this paper we shall describe {\tt EXCALIBUR}, our event
generator to compute {\em all}
4-fermion processes in $\epl \emn$ collisions, including
QED initial state corrections and QCD diagrams. The general
structure of the code is flexible enough to deal with physics at the
energy scales from 100 GeV to 1 TeV.

\section{Theory and general features}
There are two sources of complications. First af all one has to
generate and compute all possible Feynman diagrams
contributing to any given final state. Then the
Monte Carlo integration has to be performed.

As explained in ref. \cite{np} the former problem can be efficiently
solved by using spinorial helicity techniques.
The amplitudes receive contributions from Abelian and
non\--A\-be\-li\-an graphs, with two distinct topological structures
(see fig. 1).
In these so-called {\em generic\/} diagrams, all particles are assumed
to be outgoing: assigning two fermion legs to be the initial-state
fermions (by crossing), the actual Feynman diagrams are generated.
The particles and antiparticles can each be assigned in six ways to
the external lines (in principle). This gives 36 possible permutations.
The Abelian diagrams are built by selecting, for each permutation, only
those cases in which the exchanged bosons, that may be
$W^+$, $W^-$, $Z$ or $\gamma$, give rise to existing and charge
conserving vertices. In the non\--A\-be\-li\-an diagrams,
two of the vector bosons
are fixed to be $W^+$ and $W^-$, and the third one can be $Z$ or
$\gamma$.

\noindent This procedure gives, for the Abelian graphs,
a maximum of 144 different diagrams, and at most 8 for the
non\--A\-be\-li\-an diagrams.

\bfig(300,220)
\sof(20,20)
\flin{40,0}{0,20} \flin{0,20}{40,40}
\flin{40,60}{0,80} \flin{0,80}{0,120} \flin{0,120}{40,140}
\flin{40,160}{0,180} \flin{0,180}{40,200}
\wlin{0,20}{0,80} \wlin{0,120}{0,180}
\Text(45,0)[lc]{$p_6,\sigma$} \Text(45,40)[lc]{$p_5,\sigma$}
\Text(45,60)[lc]{$p_4,\rho$} \Text(45,140)[lc]{$p_3,\rho$}
\Text(45,160)[lc]{$p_2,\lambda$} \Text(45,200)[lc]{$p_1,\lambda$}
\Text(-5,50)[rc]{$V_1$} \Text(-5,150)[rc]{$V_2$}
\sof(200,40)
\flin{40,0}{0,20} \flin{0,20}{40,40}
\wlin{0,20}{0,80} \wlin{0,80}{40,80} \wlin{0,80}{0,140}
\flin{80,60}{40,80} \flin{40,80}{80,100}
\flin{40,120}{0,140} \flin{0,140}{40,160}
\Text(45,0)[lc]{$p_6,\sigma$}
\Text(45,40)[lc]{$p_5,\sigma$}
\Text(85,60)[lc]{$p_4,\rho$}
\Text(85,100)[lc]{$p_3,\rho$}
\Text(45,120)[lc]{$p_2,\lambda$}
\Text(45,160)[lc]{$p_1,\lambda$}
\Text(-5,50)[rc]{$W^-$}
\Text(-5,110)[rc]{$W^+$}
\Text(20,75)[tc]{$V$}
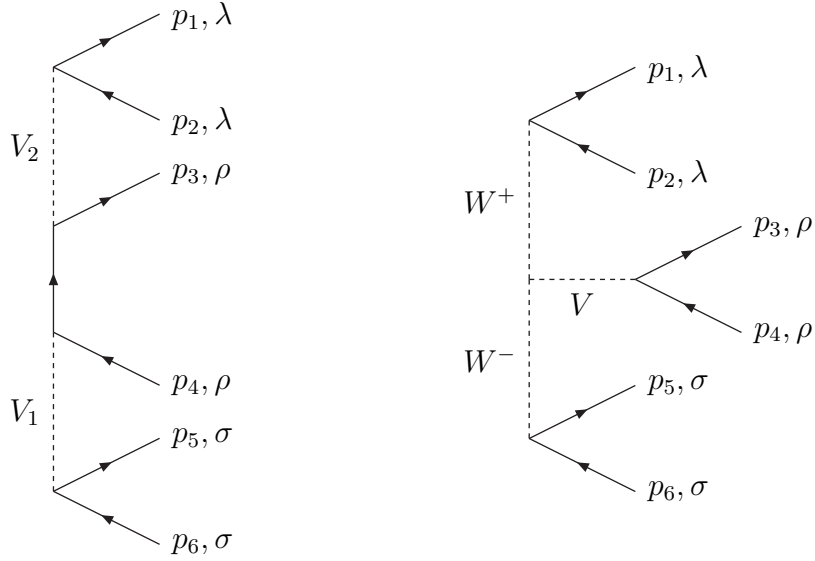
\efig{Figure 1: generic diagrams for four-fermion production.
The fermion momenta and helicities, and the bosons are indicated.
The bosons $V_{1,2}$ can be either $Z$, $W^{\pm}$, or $\gamma$;
$V$ can be either $Z$ or $\gamma$.}

\noindent The spinorial structure of each diagram can always be written in such
a way that a particular combination of axial and vector couplings
factorizes
for a given helicity assignment. For example, if $a_i$ and $v_i$
are the axial and vector couplings in the vertices of the abelian
diagram of fig. 1, the following equation holds
\bqa
\label{split}
& & \ubar_{\lambda}(1)\gamma^{\mu}(v_1+a_1\gamma_5)u_{\lambda}(2)
    \times \\
& & \ubar_{\rho}(3)\gamma_{\mu}(v_2+a_2\gamma_5)
(\rlap/p_1+\rlap/p_2+\rlap/p_3)
\gamma_{\nu}(v_3+a_3\gamma_5)u_{\rho}(4)\times \nl
& & \ubar_{\sigma}(5)\gamma^{\nu}(v_4+a_4\gamma_5)u_{\sigma}(6)=
\sum_{\alpha, \beta, \tau= \pm} {\cal P}(\lambda\beta,\rho\alpha,
\sigma\tau)\, A(\lambda,\rho,\sigma;1,2,3,4,5,6) \nonumber
\eqa
where
\bqa
\label{vertex}
\lefteqn{{\cal P}(\lambda\beta,\rho\alpha,\sigma\tau)= P_{\lambda\beta}
P_{\rho\alpha}P_{\sigma\tau}\,V^\beta_1V^\alpha_2V^\tau_3} \\
\lefteqn{P_{\lambda\beta}= {1 \over 2}(1+\lambda\beta)} \nl
\lefteqn{V^\pm_1= v_1 \pm a_1} \nl
\lefteqn{V^\pm_2= (v_2 \pm a_2)(v_3 \pm a_3)} \nl
\lefteqn{V^\pm_3= v_4 \pm a_4} \nl
\lefteqn{A(\lambda,\rho,\sigma;1,2,3,4,5,6) =}\nl
& & \ubar_{\lambda}(1)\gamma^{\mu}u_{\lambda}(2)\times \nl
& & \ubar_{\rho}(3)\gamma_{\mu}(\rlap/p_1+\rlap/p_2+\rlap/p_3)
\gamma_{\nu}u_{\rho}(4)\times \nl
& & \ubar_{\sigma}(5)\gamma^{\nu}u_{\sigma}(6)\;\;.\nonumber
\eqa
Here we have disregarded the particle/antiparticle distinction
since it is already implied by the assignment of
the external momenta.
The helicity labels $\lambda,\rho,\sigma = \pm$ determine the helicity
of both external legs on a given fermion line. Using the
Weyl-van der Waerden formalism for helicity amplitudes \cite{spinors8}
(or, equivalently, the Dirac formalism of \cite{spinors7}),
the expression $A$ can easily be calculated \cite{np}.
It turns out that, for each permutation of the fermion momenta,
all helicity combinations can be computed using only four independent
complex functions.

\noindent The numerator in the non\--A\-be\-li\-an diagrams can also be written
in
terms of the function $A$:
\bqa
\lefteqn{\ubar_{\lambda}(1)\gamma_{\alpha}u_{\lambda}(2)
       \;\ubar_{\rho}(3)\gamma_{\mu}u_{\rho}(4)
       \;\ubar_{\sigma}(5)\gamma_{\nu}u_{\sigma}(6)}\nl
\lefteqn{\times\;2\,
 \left\{g^{\mu\alpha}(p_1+p_2)^{\nu}
       +g^{\alpha\nu}(p_5+p_6)^{\mu}
       +g^{\nu\mu}(p_3+p_4)^{\alpha}\right\}}\nl
& = & A(\lambda,\rho,\sigma;1,2,3,4,5,6) -
      A(\sigma,\rho,\lambda;5,6,3,4,1,2)\;\;.
\eqa
Thus, for massless fermions, every helicity amplitude consists of a sum of
very systematic, and relatively compact, expressions.

When four quarks are present in the final state, one has to add
the concomitant QCD production channels, and also the production of a quark
pair and two gluons, since both types of final states will appear as jets.
The former contribution, which we call {\em interfering} QCD background,
is easily implemented  once all electroweak diagrams
have been computed as shown before. In fact, it is enough to add gluons
wherever
photons connect quark lines
\cite{qcd} (of course the correct QCD coupling and colour structure should
be taken into account).  Finally, the latter process
can be efficiently computed using the recursion relations of ref.
\cite{giele}.
Since it does not interfere with the other diagrams, we have written a
separate event generator to get this contribution \cite{qqgg}.
For the sake of brevity we do not describe it here. However, we point out
that, as for the Monte Carlo integration, it has been built following
exactly the same strategy used in {\tt EXCALIBUR}.

The problem of the integration over the final fermion momenta can be
solved using a {\em multichannel} approach \cite{multi10,np}.
If $f(\vec \Phi)$ denotes the matrix element squared
and $d\vec \Phi$ the 8-dimensional massless phase space integration
element, one has to compute
\bqa
\sigma= \int f(\vec\Phi)\;d\vec\Phi\;\theta(cuts)\;\;,
\eqa
where $\theta(cuts)$ stands for any kind of experimental cut, that,
in a Monte Carlo approach, is implemented by simply putting
$f(\vec\Phi)=0$ in the unwanted region of the phase space.

\noindent In order to reduce the variance of the integrand, and therefore the
Monte Carlo error, it is convenient to introduce an analytically
integrable function $g(\vec \Phi)$, called the {\em local density\/},
that exhibits approximately the same peaking
behaviour of $f(\vec\Phi)$ and is {\em unitary\/}, that is,
a normalized probability density:
\bqa
\int g(\vec\Phi)\;d\vec\Phi= 1\;\;.
\eqa
By multiplying and dividing the integrand by $g(\vec \Phi)$, the cross
section can be rewritten as follows
\bqa
\sigma= \int w(\vec\Phi(\vec\rho))\;d\vec\rho\;\theta(cuts)
\eqa
where the new integrand
\bqa
w(\vec\Phi(\vec\rho)) &=&\frac{f(\vec\Phi)}{g(\vec\Phi)}
\label{weight}
\eqa
is a smoother function of the new set of variables $\{\rho_i\}$ defined
by
\bqa
d\vec\rho = g(\vec\Phi)d\vec\Phi \nonumber \\
0<\rho_i<1\;\;
\eqa
so that the variance of $w(\vec\rho)$ is smaller than the variance of
$f(\vec\Phi)$.

\noindent When the peaking structure of the matrix element squared
is very rich one set of new integration variables
$\{ \rho_i\}$ can only describe well a limited number of peaks.
Therefore a {\em multichannel} approach is required in which
\bq
g(\vec \Phi) =
\_{i= 1}^N \alpha_i\;g_i(\vec \Phi)
\;\;,\;\;
\sum_{i= 1}^N \alpha_i = 1\;\;,\;\;
\int g_i(\vec \Phi) d \vec \Phi = 1\;\;,
\eq
and where every $g_i(\vec \Phi)$ describes a particular peaking structure
of $f(\vec\Phi)$. Note that the conditions on the $\alpha_i$ and
$g_i(\Phi)$ ensure unitarity of the algorithm, {\it i.e.\/}
probability is explicitly conserved at each step of the
algorithm, without additional normalization factors at any stage.

\noindent The numbers $\alpha_i$ are called {\em a-priori weights} and,
although their numerical values are in principle unimportant,
they can be used, in practice, to reduce the Monte Carlo error
\cite{cpc2}.

\noindent In {\tt EXCALIBUR} we have dealt with the problem of the construction
of
the $g_i(\vec\Phi)$ in a very modular and systematic way.
Firstly, we have singled out all possible
kinematical diagrams occurring in a four-fermion final state (see fig. 2).
They are pictures, inspired by the Feynman diagrams, which represent
the various peaking structures of the matrix element and
indicate which variables are most appropriate to a given
$g_i(\vec\Phi)$. The explanation of the pictures will be given in
section 3.1. Secondly, we have written all
building blocks (that is subroutines) necessary for the calculation.
Finally, we have put them together to form the $g_i(\vec\Phi)$.

QED corrections are implemented using the structure-function
formalism \cite{qedsf,qed}. Each of the incoming fermions is assumed
to have its energy
degraded by the emission of photons parallel to the beam.
For the energy distribution of the fermion after radiation
we take a structure function $\Phi$ that incorporates
the leading log $\order{}$ and $\order{2}$ initial
state radiation with exponentiation of the remaining soft-photon effects.
Its expression can be found in \cite{qed}.
Our model for the total radiative cross
section is then
\bq
\sigma(s) =  \int\limits_0^1\int\limits_0^1\,
dx_1\,dx_2\,\Phi(x_1)\,\Phi(x_2)\,\sigma_0(x_1x_2s)
\label{cross}
\eq
where $\sigma_0$ is the non-radiative cross section and $x_1,\, x_2$
represent the energy content of the incoming fermions after
radiative emission.
This provides an adequate description of the
leading QED effects.
\section{Program Structure}
We shall now describe in some detail the salient features and strategies
adopted in {\tt EXCALIBUR}.
The Program consists of two parts: the evaluation of the matrix element
and the event generation.
Both steps require an initialization, according to the chosen final
state. Roughly speaking it means that the Feynman
diagrams and the kinematical channels have to be built.
This initialization is done in {\tt SUBROUTINE SETPRO}, the matrix
element is evaluated in {\tt SUBROUTINE DIAGA} and
{\tt SUBROUTINE MATRIX}, while nearly all the rest is devoted to the
event generation and Monte Carlo integration.
\subsection{Subroutine SETPRO}
We already described the algorithm used to construct
the Feynman diagrams through a big do loop over all 36 permutations
of the six fermion momenta. In {\tt SUBROUTINE SETPRO} the variable
{\tt KPERM(1:6,1:36)} explicitly contains all these permutations, and
{\tt IPHASE(1:36)} the corresponding relative phase.
The constructed Abelian (non Abelian) diagrams are
stored in {\tt JJ(1:16,NDAB)} ({\tt JN(1:16,NNAB)}), where the first index
contains
information about the particles involved in the process, the vector
boson propagators and the momenta assignment, and the second one
enumerates each diagram.
In the output each constructed diagram is printed out together
with its list number. For particular studies or checks, we
give the possibility to switch off diagrams. This can be achieved
by putting the variables {\tt KA0(I)}=0 ({\tt K10(J)}=0),
for the corresponding unwanted Abelian (non Abelian) diagrams
({\tt I=1:NDAB, J=1:NNAB}).
{\tt SUBROUTINE SETPRO} also contains the the input parameters
of the program. They are $\alpha$ ({\tt ALPHA}), $\alpha_s$
({\tt ALS}, relevant for 4 quarks final states), $M_Z$ ({\tt ZM}),
$M_W$ ({\tt WM}), $\sin^2 \theta_W$ ({\tt STH2}), $\Gamma_W$
({\tt WW}) and  $\Gamma_Z$ ({\tt WZ}). The statistical factor
{\tt STATFAC} and the colour factor {\tt FCOL} are evaluated according
to the chosen final state. Furthermore, the coupling
combinations $V^\pm_i$ of eq. \ref{vertex} (and those occurring
in the non Abelian case) are computed. It may happen that, for a
particular helicity combination, one or more of the $V^\pm_i$ are zero.
In the latter cases there is no point in computing the corresponding
function $A$ (see eq. \ref{split}).
As a result, less than four independent complex functions are
required to evaluate the spinorial part of the diagram.
In order to have a fast evaluation of the matrix element those cases
have to be excluded. This is achieved by introducing
two {\em occupation matrices} {\tt NC(1:36)} and {\tt NOC(1:36,1:4)}.
For each of the 36 permutation, {\tt NC} is set zero if
the corresponding permutation does not give any Feynman diagram, and,
if it does, {\tt NOC} indicates which complex functions are needed.
Through {\tt COMMON/AREA3/} these matrices are passed to {\tt SUBROUTINE
DIAGA}, where only those helicity combination for which
{\tt NOC} and {\tt NC} are different from zero are computed.

\noindent Two more operations are performed in {\tt SUBROUTINE SETPRO}, namely
the choice of the kinematical channels for the Monte Carlo integration
and the computation of the QCD interfering background.

\noindent We singled out a maximum number of 26 kinematical channels.
They are given in fig. 2, together with the name of the
corresponding subroutines in {\tt EXCALIBUR}, and are inspired by
all possible occurring Feynman diagrams.
Fermionic lines have an arrow, a wavy line represents a photon and
a dashed line can be either $Z$ or $W$ (this gives 26 channels).
Solid lines connect topological equivalent points. That is
a {\em t-channel} solid line means isotropic angular distributions
between the connected fermions while a {\em s-channel} solid line
stands for photon or massive vector boson propagators. Since they
only give rise to an $s$ dependent behaviour, the peaking structure
relevant for the integration over the final momenta is not affected
by them.
As an example, with those conventions it is easy to recognize that
the last channel {\tt RAMBO4} represents an isotropic 4 body decay.
If the {\em inspiring} Feynman diagrams exist,
the variables {\tt NCHA(1:26,1:48)} are set equal to one, where the
first index runs over the possible channels and the
second one labels the permutation of the final momenta.
The number 48 is explained as follows.
There are 24 permutations of the four final momenta but, for some
channel, the case where the initial state labels 1 and 2 refer
to $e^+$ and $e^-$ respectively must be distinguished from the case
where they refer to $e^-$ and $e^+$. This gives 48 possible
permutations.
Depending on the topology, there are symmetries among the final momenta
that have to be taken into account in order to have a minimum number
of kinematical channels. For example, in channel {\tt NONAB1},
permutation 3456 of the final momenta is equivalent to permutation 5634.
This {\em symmetrization} is automatically performed by the program.
When the initialization in {\tt SUBROUTINE SETPRO} is completed,
variable {\tt NCT} indicates the number of found kinematical channels.
In the output file they are printed out together with a list number
{\tt I}. An array has been introduced ({\tt NCTO}), such that
putting {\tt NCTO(I)}= 0 excludes channel with number {\tt I} ({\tt I=1:NCT})
when the Monte Carlo integration is performed. This can be used to
increase the speed of the program, by switching off those channels for
which the procedure of self-optimization (see below) gives very small
a-priori weights.

\noindent The interfering QCD background is added
as an extra contribution proportional to the ratio
${\alpha_s \over \alpha Q Q'}$ (the variable {\tt GRAP}),
for those amplitudes where a photon connects two quarks of charge
$Q$ and $Q'$.

\noindent This concludes the description of {\tt SUBROUTINE SETPRO}.
Since all possible initializations are performed there, the structure
of the rest of the program can be simple and fast.
\subsection{Subroutines DIAGA and MATRIX}
In {\tt SUBROUTINE DIAGA} numerators and denominators of all found
Feynman diagrams for which {\tt NOC} and {\tt NC} are non vanishing
are computed {\em at once} using the Weyl-van der Waerden
formalism. It means that, for each generated event,
{\tt SUBROUTINE DIAGA} is called just once and not $n$ times, where
$n$ is the total number of Feynman diagrams.
As for the computational speed, this is very important.

\noindent In {\tt SUBROUTINE MATRIX(SQUAREM)} the matrix element squared
({\tt SQUAREM}) is calculated by putting together the numerators and
denominators computed in {\tt SUBROUTINE DIAGA} and the coupling
combinations of eq. \ref{vertex} evaluated in {\tt SUBROUTINE SETPRO}.
Since computing the colour factor and QCD interfering background  in a
four quark final state with colour labels {\em i, j, l} and {\em m}
requires the part of the amplitude proportional to
$\delta_{ij}\delta_{lm}$ to be distinguished from that proportional
to $\delta_{il}\delta_{jm}$
\cite{qcd}, the constructed amplitudes in {\tt SUBROUTINE MATRIX}
take care of both contributions separately.
\subsection{Phase space generation and integration}
In the {\tt MAIN} of {\tt EXCALIBUR} the variables
{\tt XR1} and {\tt XR2}, representing the energy content $x_1$ and $x_2$
of the incoming fermions after radiative emission (eq. \ref{cross}),
are generated. Then, the initial configuration of the momenta {\em
in the center of mass frame of the event after ISR} is set
calling {\tt SUBROUTINE MOMSET} and the cuts imposed on the momenta
in the Lab frame are rewritten in terms of cuts in the center of mass
frame. The kinematical channels are called using
{\tt SUBROUTINE ADDRESS(LFLAG,NC,NN,DJ)}.
When {\tt LFLAG} is set 0, the channel number {\tt NC},
with the momenta permutation labelled by {\tt NN}, is used for
generating the momenta and computing the local density {\tt DJ}.
If {\tt LFLAG}= 1 the actual momenta configuration is used
to compute {\tt DJ}.
The choice of the channel to use is performed, as in ref.
\cite{multi10}, on the basis of the actual values of the a-priori
weights $\alpha_i$ by defining the cumulative numbers
$\beta_i= \alpha_1 + ... + \alpha_i$, taking a random number
uniformly distributed between 0 and 1 and choosing channel $i$ if
$\beta_{i-1}<z<\beta_i$.

\noindent In {\tt SUBROUTINE MOMARRAY} the generated four-momenta are
put in a big array {\tt PM(0:4,0:900)} and stored in

\vspace{0.3cm}

{\tt COMMON/MOMENTA/ROOTS,XR1,XR2,PM(0:4,0:900)}

\vspace{0.3cm}

\noindent ({\tt ROOTS} is the center of mass energy of the event).
The first index refers to the component of the momenta (0 represents
the energy and 4 is the four momentum squared). As for the second index,
the following self-explanatory conventions are used:

\vspace{0.3cm}

{\tt PM(I,34) $\equiv$ PM(I,43)= PM(I,3)+PM(I,4)}~~~etc.

{\tt PM(I,643) $\equiv$ PM(I,346) $\equiv$ ... =
          PM(I,6)+PM(I,4)+PM(I,3)}~~~etc.

\vspace{0.3cm}

\noindent Besides, but only if the first digit refers to an incoming momentum
(notice the correspondence $7 \to 1$, $8 \to 2$)

\vspace{0.3cm}

{\tt PM(I,734) $\equiv$ PM(I,1)-PM(I,3)-PM(I,4)}~~~etc.

{\tt PM(I,851) $\equiv$ PM(I,2)-PM(I,5)-PM(I,1)}~~~etc.

\vspace{0.3cm}

\noindent Each channel is constructed in a very modular way
by putting together basic subroutines that describe different parts of
its peaking structure.
In ref. \cite{np} an example of the construction of
channels {\tt BREMB2} and {\tt CONVER2} of fig. 2 is given.
There are 10 of these basic subroutines.
They are the {\em building blocks} of the whole
generation procedure. For the sake of
brevity we do not list them here. They are well commented in
the program. We only notice that, in building the
kinematical channels, every t-channel exchanged massive vector boson
is always assumed to give a flat angular distribution between the initial
and the final fermion. This is done in order to avoid
proliferation in the number of channels. In our experience,
this gives a very good approximation at center of mass energies
up to 500 GeV, a good approximation at higher energies up to 1 TeV and
may cause large Monte Carlo errors at 2 Tev. Of course the Monte Carlo
program remains correct, but higher statistics runs are required.
However, adding channels to map this high energy kinematical behaviour
is trivial, because {\tt EXCALIBUR} already contains all needed
ingredients.

As far as the self-optimization of the integration is concerned,
a detailed description of the iterative algorithm implemented in
{\tt EXCALIBUR} may be found in ref. \cite{cpc2}. Here we point
out that two variables have to be chosen by the user, namely
the maximum number of iterations {\tt ISTEPMAX} (in the input list)
and the number of point {\tt NOPT} used for the self-optimization
(in the {\tt MAIN} of the program). Then, for each iteration,
{\tt NOPT}/{\tt ISTEPMAX} points (including zero-weight events)
are used to compute the a-priori weights.
We found that, with 4-5 hundred thousand points,
a good choice is {\tt NOPT=} 100,000 and {\tt ISTEPMAX=} 10.
However, when very stringent cuts are applied, the
majority of the events falls outside the allowed region, so that the
ratio {\tt NOPT}/{\tt ISTEPMAX} may be a very small number. This causes
a bad estimate of the best a-priori weights to be used.
In those cases it is convenient to either increase {\tt NOPT} or
decrease {\tt ISTEPMAX}.

In the input list one has to specify the set of {\em standard}
cuts as specified in the next section.
Any other type of cut must be implemented directly in
{\tt SUBROUTINE CUTS(LNOT)}, where

\vspace{0.3cm}

{\tt COMMON/AREA10/PM1(0:4,1:6),PM4(12:65),OMCT1(1:6,3:6)}

\vspace{0.3cm}

\noindent contains the four momenta computed {\em in the Lab frame}
({\tt PM1}), the invariant mass squared among all possible particles
pairs ({\tt PM4}) and the quantities $1-\cos \theta_{ij}$ ({\tt OMCT1}).
If the event is rejected {\tt LNOT=} 1, and the weight is put
equal to zero.

Finally, all weights (computed as in eq. \ref{weight}) are collected
using {\tt SUBROUTINE INBOOK} and the Monte Carlo results called
through {\tt SUBROUTINE OUTBOK}.

\section{Input}

The meaning of the input parameters is the following:

\vspace{0.4cm} \noindent {\tt NPROCESS(INTEGER)}

\vspace{0.2cm} \noindent The number of processes to be computed.

\vspace{0.4cm} \noindent {\tt N(INTEGER)}

\vspace{0.2cm} \noindent The number of points for the Monte Carlo
                          integration.

\vspace{0.4cm} \noindent {\tt ISTEPMAX(INTEGER)}

\vspace{0.2cm} \noindent The number of iterations for optimizing the
                         a-priori  weights.

\vspace{0.4cm} \noindent {\tt OUTPUTNAME(CHARACTER*15)}

\vspace{0.2cm} \noindent The name of the output file.

\vspace{0.4cm} \noindent {\tt KREL(INTEGER)}

\vspace{0.2cm} \noindent It selects the signals. If {\tt KREL=} 0
               all Feynman diagrams are taken into account. If
               {\tt KREL=} 1-5 only those leading to
               reactions of eqs. (\ref{wpair})-(\ref{singleznu}).

\vspace{0.4cm} \noindent {\tt LQED(INTEGER)}

\vspace{0.2cm} \noindent It includes (1) or excludes (0) ISR.

\vspace{0.4cm} \noindent {\tt ROOTSMUL(REAL*8)}

\vspace{0.2cm} \noindent The total energy of the colliding $e^+$
                         and $e^-$. All energies are in GeV.

\vspace{0.4cm} \noindent {\tt SHCUT(REAL*8)}

\vspace{0.2cm} \noindent Minimum value of the invariant mass squared of the
                        event after QED radiation.

\vspace{0.4cm} \noindent {\tt ECUT(3)(REAL*8)}

\vspace{0.2cm} \noindent Minimum energy of particle number 3.

\vspace{0.4cm} \noindent {\tt ECUT(4)(REAL*8)}

\vspace{0.2cm} \noindent Minimum energy of particle number 4.

\vspace{0.4cm} \noindent {\tt ECUT(5)(REAL*8)}

\vspace{0.2cm} \noindent Minimum energy of particle number 5.

\vspace{0.4cm} \noindent {\tt ECUT(6)(REAL*8)}

\vspace{0.2cm} \noindent Minimum energy of particle number 6.

\vspace{0.4cm} \noindent {\tt SCUT(3,4)(REAL*8)}

\vspace{0.2cm} \noindent Minimum value of $(p(3)+p(4))^2$. All invariant
masses are in ${\rm GeV}^2$.

\vspace{0.4cm} \noindent {\tt SCUT(3,5)(REAL*8)}

\vspace{0.2cm} \noindent Minimum value of $(p(3)+p(5))^2$.

\vspace{0.4cm} \noindent {\tt SCUT(3,6)(REAL*8)}

\vspace{0.2cm} \noindent Minimum value of $(p(3)+p(6))^2$.

\vspace{0.4cm} \noindent {\tt SCUT(4,5)(REAL*8)}

\vspace{0.2cm} \noindent Minimum value of $(p(4)+p(5))^2$.

\vspace{0.4cm} \noindent {\tt SCUT(4,6)(REAL*8)}

\vspace{0.2cm} \noindent Minimum value of $(p(4)+p(6))^2$.

\vspace{0.4cm} \noindent {\tt SCUT(5,6)(REAL*8)}

\vspace{0.2cm} \noindent Minimum value of $(p(5)+p(6))^2$.

\vspace{0.4cm} \noindent {\tt CMAX(1,3)(REAL*8)}

\vspace{0.2cm} \noindent Maximum value of $\cos \theta$ between
                         particle 1 and 3.

\vspace{0.4cm} \noindent {\tt CMAX(1,4)(REAL*8)}

\vspace{0.2cm} \noindent Maximum value of $\cos \theta$ between
                         particle 1 and 4.

\vspace{0.4cm} \noindent {\tt CMAX(1,5)(REAL*8)}

\vspace{0.2cm} \noindent Maximum value of $\cos \theta$ between
                         particle 1 and 5.

\vspace{0.4cm} \noindent {\tt CMAX(1,6)(REAL*8)}

\vspace{0.2cm} \noindent Maximum value of $\cos \theta$ between
                         particle 1 and 6.

\vspace{0.4cm} \noindent {\tt CMAX(2,3)(REAL*8)}

\vspace{0.2cm} \noindent Maximum value of $\cos \theta$ between
                         particle 2 and 3.

\vspace{0.4cm} \noindent {\tt CMAX(2,4)(REAL*8)}

\vspace{0.2cm} \noindent Maximum value of $\cos \theta$ between
                         particle 2 and 4.

\vspace{0.4cm} \noindent {\tt CMAX(2,5)(REAL*8)}

\vspace{0.2cm} \noindent Maximum value of $\cos \theta$ between
                         particle 2 and 5.

\vspace{0.4cm} \noindent {\tt CMAX(2,6)(REAL*8)}

\vspace{0.2cm} \noindent Maximum value of $\cos \theta$ between
                         particle 2 and 6.

\vspace{0.4cm} \noindent {\tt CMAX(3,4)(REAL*8)}

\vspace{0.2cm} \noindent Maximum value of $\cos \theta$ between
                         particle 3 and 4.

\vspace{0.4cm} \noindent {\tt CMAX(3,5)(REAL*8)}

\vspace{0.2cm} \noindent Maximum value of $\cos \theta$ between
                         particle 3 and 5.

\vspace{0.4cm} \noindent {\tt CMAX(3,6)(REAL*8)}

\vspace{0.2cm} \noindent Maximum value of $\cos \theta$ between
                         particle 3 and 6.

\vspace{0.4cm} \noindent {\tt CMAX(4,5)(REAL*8)}

\vspace{0.2cm} \noindent Maximum value of $\cos \theta$ between
                         particle 4 and 5.

\vspace{0.4cm} \noindent {\tt CMAX(4,6)(REAL*8)}

\vspace{0.2cm} \noindent Maximum value of $\cos \theta$ between
                         particle 4 and 6.

\vspace{0.4cm} \noindent {\tt CMAX(5,6)(REAL*8)}

\vspace{0.2cm} \noindent Maximum value of $\cos \theta$ between
                         particle 5 and 6.

\vspace{0.4cm} \noindent {\tt PAR(3)(CHARCTER*8)}

\vspace{0.2cm} \noindent Produced fermion with label 3 (to be chosen
                         among {\tt 'EL'}, {\tt 'NE'}, {\tt 'MU'},
                               {\tt 'NM'}, {\tt 'TA'}, {\tt 'NT'},
                               {\tt 'DQ'}, {\tt 'UQ'}, {\tt 'SQ'},
                               {\tt 'CQ'}, {\tt 'BQ'}, {\tt 'TQ'}).

\vspace{0.4cm} \noindent {\tt PAR(4)(CHARCTER*8)}

\vspace{0.2cm} \noindent Produced antifermion with label 4.

\vspace{0.4cm} \noindent {\tt PAR(5)(CHARCTER*8)}

\vspace{0.2cm} \noindent Produced fermion with label 5.

\vspace{0.4cm} \noindent {\tt PAR(6)(CHARCTER*8)}

\vspace{0.2cm} \noindent Produced antifermion with label 6.

\section{Test Run Output}
To conclude our description, we give an example of a
typical calculation that can be performed with {\tt EXCALIBUR}.
One should be able
to reproduce this output within the estimated Monte Carlo error
(small differences may occur because the quasi-random number generator
used in the program is not strictly portable). Using an input file
as follows

\begin{verbatim}
1             number of energy points
250000        number of Monte Carlo points
10            number of iterations in a.p.weights optimization
output        output program name
0             krel (signal:  0,1,2,3,4,5)
1             lqed (0 or 1)
190.d0        total energy (GeV)
0.d0          cut on reduced inv. mass squared after ISR
0.d0          ecut_3
0.d0          ecut_4
20.d0         ecut_5
20.d0         ecut_6
0.d0          scut_34
0.d0          scut_35
0.d0          scut_36
0.d0          scut_45
0.d0          scut_46
100.d0        scut_56
1.d0          cmax_13
1.d0          cmax_14
0.9d0         cmax_15
0.9d0         cmax_16
1.d0          cmax_23
1.d0          cmax_24
0.9d0         cmax_25
0.9d0         cmax_26
1.d0          cmax_34
1.d0          cmax_35
1.d0          cmax_36
1.d0          cmax_45
1.d0          cmax_46
0.9d0         cmax_56
mu            produced fermion     (3)
nm            produced antifermion (4)
uq            produced fermion     (5)
dq            produced antifermion (6)
\end{verbatim}
and the values {\tt ALPHA= 1./128.}, {\tt ZM= 91.16},
{\tt WM= 80.22}, {\tt STH2= 0.226 }, {\tt WW= 2.03}, {\tt WZ= 2.53}
we get the following output file
\begin{verbatim}
                    output

All Feynman diagrams

sqrt(s)  =     .190000D+03
n_points =   250000
istepmax =       10

energy cuts with ecut_3  = .0
                 ecut_4  = .0
                 ecut_5  = 20.0
                 ecut_6  = 20.0

cut on        s*x1r*x2r  = .0

mass cuts with  scut_34  = .0
                scut_35  = .0
                scut_36  = .0
                scut_45  = .0
                scut_46  = .0
                scut_56  = 100.0

angle cuts with cmax_13  = 1.0
                cmax_14  = 1.0
                cmax_15  = .9
                cmax_16  = .9
                cmax_23  = 1.0
                cmax_24  = 1.0
                cmax_25  = .9
                cmax_26  = .9
                cmax_34  = 1.0
                cmax_35  = 1.0
                cmax_36  = 1.0
                cmax_45  = 1.0
                cmax_46  = 1.0
                cmax_56  = .9

I.S.R. INCLUDED

s^2_thet =     .226000D+00
Z-mass   =     .911600D+02
Z-width  =     .253000D+01
W-mass   =     .802200D+02
W-width  =     .203000D+01
1/alpha  =     .128000D+03
alpha_s  =     .103000D+00

process:  antiel(1) el(2) ---> mu(3) antinm(4) uq(5) antidq(6)

          abelian diagrams                              phase

1:  [el(1),el(2)] Z [mu(3),mu,nm(4)] W [uq(5),dq(6)]    ph=  1
2:  [el(1),el(2)] G [mu(3),mu,nm(4)] W [uq(5),dq(6)]    ph=  1
3:  [el(1),el(2)] Z [uq(5),uq,dq(6)] W [mu(3),nm(4)]    ph=  1
4:  [el(1),el(2)] G [uq(5),uq,dq(6)] W [mu(3),nm(4)]    ph=  1
5:  [mu(3),nm(4)] W [uq(5),dq,dq(6)] Z [el(1),el(2)]    ph=  1
6:  [mu(3),nm(4)] W [uq(5),dq,dq(6)] G [el(1),el(2)]    ph=  1
7:  [uq(5),dq(6)] W [el(1),ne,el(2)] W [mu(3),nm(4)]    ph=  1
8:  [uq(5),dq(6)] W [mu(3),nm,nm(4)] Z [el(1),el(2)]    ph=  1

          non abelian diagrams                          phase

1:  [uq(5),dq(6)] [el(1),el(2)] [mu(3),nm(4)] (WZW)     ph=  1
2:  [uq(5),dq(6)] [el(1),el(2)] [mu(3),nm(4)] (WGW)     ph=  1

          kinematical diagrams

    channel                     permutation

1:  annihi2(wm)                 1 2 3 4 5 6
2:  annihi2(wm)                 1 2 4 3 5 6
3:  annihi2(wm)                 1 2 5 6 3 4
4:  annihi2(wm)                 1 2 6 5 3 4
5:  conver3(wm)                 1 2 5 6 3 4
6:  nonab1(wm)                  1 2 3 4 5 6
7:  rambo4                      1 2 3 4 5 6

 ********** weights analysis **********

 *** variable number  1 ************
sum(w**0)       .250000D+06,  sum(w**1)         .135757D+06
sum(w**2)       .254349D+06,  sum(w**3)         .803758D+06
sum(w**4)       .434411D+07
maximum         .221013D+02,  max.in buffer     .148670D+02
no.weights=0          59062,  no.weights<0                0
estimator x:     .543028D+00
estimator y:     .289008D-05
estimator z:     .730267D-15
average estimate :    .543028D+00
              +\-     .170002D-02
variance estimate:    .289008D-05
              +\-     .270235D-07
efficiency for all weights     :    2.457 %
efficiency for nonzero weights :    3.217 %
overshoot factor of histogram  :    1.487
the distribution of the nonzero weights:
50, log scale;  entries under,inside,over: 0  190928      10

  .1487E+01   .1631E+06     i******************************i
  .2973E+01   .2157E+05     i*************************     i
  .4460E+01   .4936E+04     i**********************        i
  .5947E+01   .1089E+04     i******************            i
  .7433E+01   .1470E+03     i*************                 i
  .8920E+01   .3600E+02     i*********                     i
  .1041E+02   .1600E+02     i*******                       i
  .1189E+02   .1100E+02     i******                        i
  .1338E+02   .7000E+01     i*****                         i
  .1487E+02   .5000E+01     i*****                         i

differences in the computation
of the a-priori weights:

    diff( 1)=  1.95407687917905
    diff( 2)=  .891414805616945
    diff( 3)=  .7677390572607608
    diff( 4)=  .6967978248575471
    diff( 5)=  .5755243287941117
    diff( 6)=  .7280986181406474
    diff( 7)=  .5755417220460467
    diff( 8)=  .5925286045906328
    diff( 9)=  .5632310939026423
    diff( 10)=  .598279075660049
    diff( 11)=  .5921705560137981
    diff( 12)=  .5181929881140225

  a-priori weights:

  1 :      .899493D-03
  2 :      .165631D-03
  3 :      .114732D-03
  4 :      .460900D-03
  5 :      .881137D+00
  6 :      .117221D+00
  7 :      .817789D-06
\end{verbatim}

\noindent After information about input parameters and imposed cuts, the
program prints out the used Feynman diagrams and kinematical channels.
Then, the analysis of the weights giving the Monte Carlo
estimate of the cross section (variable number 1) follows.
In particular various sums of the weights to powers 0-4 are given as
well as the maximum weight and that one in the buffer (that is in the
interval of values used in the histogram that shows the weight
distribution).
The quantity $x$ is the estimator of the average of the
distribution defined, for N weights $w_i$, as
\bq
{\sum_i w_i \over N}\;,
\eq
$y$ is the estimator of the variance
\bq
{1 \over N} \left[{\sum_i w^2_i \over N}- {(\sum_i w_i)^2- \sum_i w^2_i
\over N(N-1)} \right]\;,
\eq
and $z$ is an estimator for the variance of the variance, so that
the error on the average and variance estimates are $\sqrt{y}$
and $\sqrt{z}$ respectively.
In the example $\sigma=$ .5430 $\pm$ .00170 pb.
As usual the efficiency is defined as ${<w> \over max(w)}$ and the
overshoot factor is the ratio between the maximum weight and
the maximum weight written in the buffer.
In the histogram the 190928 non zero weights are
displayed according to their abundance in bins.
Finally, the variables $D$ of ref.
\cite{cpc2} (that measure, at each step in the
optimization procedure, how well the actual set of a-priori weights
approximates the behaviour of the optimal set) are printed out,
together with the found best set of a-priori weights.

\clearpage
\begin{center}
\begin{picture}(390,0)
\sof(0,0)
\flin{15,20}{0,0} \flin{0,40}{15,20} \lin{15,20}{30,20}
\flin{45,0}{30,20} \flin{30,20}{45,40}
\glin{37.5,10}{30,-30}{6} \flin{30,-30}{45,-10}
\flin{45,-50}{30,-30}
\Text(0,0)[r]{\small 2} \Text(0,40)[r]{\small 1}
\Text(45,40)[l]{\small 3} \Text(45,0)[l]{\small 4}
\Text(45,-10)[l]{\small 5} \Text(45,-50)[l]{\small 6}
\Text(18,-65)[t]{\tt ANNIHI1}
\sof(80,0)
\flin{15,20}{0,0} \flin{0,40}{15,20} \lin{15,20}{30,20}
\flin{45,0}{30,20} \flin{30,20}{45,40}
\wlin{37.5,10}{30,-30} \flin{30,-30}{45,-10}
\flin{45,-50}{30,-30}
\Text(0,0)[r]{\small 2} \Text(0,40)[r]{\small 1}
\Text(45,40)[l]{\small 3} \Text(45,0)[l]{\small 4}
\Text(45,-10)[l]{\small 5} \Text(45,-50)[l]{\small 6}
\Text(18,-65)[t]{\tt ANNIHI2}
\sof(160,0)
\flin{0,40}{11,40} \lin{11,40}{33,40} \flin{33,40}{45,40}
\glin{22,40}{22,0}{6}
\flin{11,0}{0,0} \lin{33,0}{11,0} \flin{45,0}{33,0}
\glin{33,0}{30,-30}{6} \flin{30,-30}{45,-10}
\flin{45,-50}{30,-30}
\Text(0,0)[r]{\small 2} \Text(0,40)[r]{\small 1}
\Text(45,40)[l]{\small 3} \Text(45,0)[l]{\small 4}
\Text(45,-10)[l]{\small 5} \Text(45,-50)[l]{\small 6}
\Text(18,-65)[t]{\tt BREMB1}
\sof(240,0)
\flin{0,40}{11,40} \lin{11,40}{33,40} \flin{33,40}{45,40}
\glin{22,40}{22,0}{6}
\flin{11,0}{0,0} \lin{33,0}{11,0} \flin{45,0}{33,0}
\wlin{33,0}{30,-30} \flin{30,-30}{45,-10}
\flin{45,-50}{30,-30}
\Text(0,0)[r]{\small 2} \Text(0,40)[r]{\small 1}
\Text(45,40)[l]{\small 3} \Text(45,0)[l]{\small 4}
\Text(45,-10)[l]{\small 5} \Text(45,-50)[l]{\small 6}
\Text(18,-65)[t]{\tt BREMB2}
\sof(320,0)
\flin{0,40}{11,40} \lin{11,40}{33,40} \flin{33,40}{45,40}
\glin{22,40}{22,0}{6}
\flin{11,0}{0,0} \lin{33,0}{11,0} \flin{45,0}{33,0}
\glin{11,0}{8,-30}{6} \flin{8,-30}{23,-10}
\flin{23,-50}{8,-30}
\Text(0,0)[r]{\small 2} \Text(0,40)[r]{\small 1}
\Text(45,40)[l]{\small 3} \Text(45,0)[l]{\small 4}
\Text(23,-10)[l]{\small 5} \Text(23,-50)[l]{\small 6}
\Text(18,-65)[t]{\tt BREMF1}
\sof(0,-140)
\flin{0,40}{11,40} \lin{11,40}{33,40} \flin{33,40}{45,40}
\glin{22,40}{22,0}{6}
\flin{11,0}{0,0} \lin{33,0}{11,0} \flin{45,0}{33,0}
\wlin{11,0}{8,-30} \flin{8,-30}{23,-10}
\flin{23,-50}{8,-30}
\Text(0,0)[r]{\small 2} \Text(0,40)[r]{\small 1}
\Text(45,40)[l]{\small 3} \Text(45,0)[l]{\small 4}
\Text(23,-10)[l]{\small 5} \Text(23,-50)[l]{\small 6}
\Text(18,-65)[t]{\tt BREMF2}
\sof(80,-140)
\flin{0,40}{11,40} \lin{11,40}{33,40} \flin{33,40}{45,40}
\lin{22,40}{22,0}
\flin{11,0}{0,0} \lin{33,0}{11,0} \flin{45,0}{33,0}
\glin{11,0}{8,-30}{6} \flin{8,-30}{23,-10}
\flin{23,-50}{8,-30}
\Text(0,0)[r]{\small 2} \Text(0,40)[r]{\small 1}
\Text(45,40)[l]{\small 3} \Text(45,0)[l]{\small 4}
\Text(23,-10)[l]{\small 5} \Text(23,-50)[l]{\small 6}
\Text(18,-65)[t]{\tt BREMF3}
\sof(160,-140)
\flin{0,40}{11,40} \lin{11,40}{33,40} \flin{33,40}{45,40}
\lin{22,40}{22,0}
\flin{11,0}{0,0} \lin{33,0}{11,0} \flin{45,0}{33,0}
\wlin{11,0}{8,-30} \flin{8,-30}{23,-10}
\flin{23,-50}{8,-30}
\Text(0,0)[r]{\small 2} \Text(0,40)[r]{\small 1}
\Text(45,40)[l]{\small 3} \Text(45,0)[l]{\small 4}
\Text(23,-10)[l]{\small 5} \Text(23,-50)[l]{\small 6}
\Text(18,-65)[t]{\tt BREMF4}
\sof(240,-140)
\flin{0,40}{15,20} \flin{15,20}{15,-30} \flin{15,-30}{0,-50}
\glin{15,20}{30,20}{3}
\glin{15,-30}{30,-30}{3}
\flin{45,0}{30,20} \flin{30,20}{45,40}
\flin{30,-30}{45,-10} \flin{45,-50}{30,-30}
\Text(0,-50)[r]{\small 2} \Text(0,40)[r]{\small 1}
\Text(45,40)[l]{\small 3} \Text(45,0)[l]{\small 4}
\Text(45,-10)[l]{\small 5} \Text(45,-50)[l]{\small 6}
\Text(18,-65)[t]{\tt CONVER1}
\sof(320,-140)
\flin{0,40}{15,20} \flin{15,20}{15,-30} \flin{15,-30}{0,-50}
\glin{15,20}{30,20}{3}
\wlin{15,-30}{30,-30}
\flin{45,0}{30,20} \flin{30,20}{45,40}
\flin{30,-30}{45,-10} \flin{45,-50}{30,-30}
\Text(0,-50)[r]{\small 2} \Text(0,40)[r]{\small 1}
\Text(45,40)[l]{\small 3} \Text(45,0)[l]{\small 4}
\Text(45,-10)[l]{\small 5} \Text(45,-50)[l]{\small 6}
\Text(18,-65)[t]{\tt CONVER2}
\sof(0,-280)
\flin{0,40}{15,20} \flin{15,20}{15,-30} \flin{15,-30}{0,-50}
\wlin{15,20}{30,20}
\wlin{15,-30}{30,-30}
\flin{45,0}{30,20} \flin{30,20}{45,40}
\flin{30,-30}{45,-10} \flin{45,-50}{30,-30}
\Text(0,-50)[r]{\small 2} \Text(0,40)[r]{\small 1}
\Text(45,40)[l]{\small 3} \Text(45,0)[l]{\small 4}
\Text(45,-10)[l]{\small 5} \Text(45,-50)[l]{\small 6}
\Text(18,-65)[t]{\tt CONVER3}
\sof(80,-280)
\flin{0,40}{11,40} \lin{11,40}{33,40} \flin{33,40}{45,40}
\glin{22,40}{22,10}{4}
\flin{22,10}{45,10} \flin{22,-20}{22,10} \flin{45,-20}{22,-20}
\glin{22,-50}{22,-20}{4}
\flin{11,-50}{0,-50} \lin{33,-50}{11,-50} \flin{45,-50}{33,-50}
\Text(0,-50)[r]{\small 2} \Text(0,40)[r]{\small 1}
\Text(45,40)[l]{\small 3} \Text(45,10)[l]{\small 4}
\Text(45,-20)[l]{\small 5} \Text(45,-50)[l]{\small 6}
\Text(18,-65)[t]{\tt MULTI1}
\sof(160,-280)
\flin{0,40}{11,40} \lin{11,40}{33,40} \flin{33,40}{45,40}
\glin{22,40}{22,10}{4}
\flin{22,10}{45,10} \flin{22,-20}{22,10} \flin{45,-20}{22,-20}
\lin{22,-50}{22,-20}
\flin{11,-50}{0,-50} \lin{33,-50}{11,-50} \flin{45,-50}{33,-50}
\Text(0,-50)[r]{\small 2} \Text(0,40)[r]{\small 1}
\Text(45,40)[l]{\small 3} \Text(45,10)[l]{\small 4}
\Text(45,-20)[l]{\small 5} \Text(45,-50)[l]{\small 6}
\Text(18,-65)[t]{\tt MULTI2}
\sof(240,-280)
\flin{0,40}{11,40} \lin{11,40}{33,40} \flin{33,40}{45,40}
\lin{22,40}{22,10}
\flin{22,10}{45,10} \flin{22,-20}{22,10} \flin{45,-20}{22,-20}
\lin{22,-50}{22,-20}
\flin{11,-50}{0,-50} \lin{33,-50}{11,-50} \flin{45,-50}{33,-50}
\Text(0,-50)[r]{\small 2} \Text(0,40)[r]{\small 1}
\Text(45,40)[l]{\small 3} \Text(45,10)[l]{\small 4}
\Text(45,-20)[l]{\small 5} \Text(45,-50)[l]{\small 6}
\Text(18,-65)[t]{\tt MULTI3}
\sof(320,-280)
\flin{0,15}{15,-5} \flin{15,-5}{0,-25}
\lin{15,-5}{30,-5}
\wlin{30,20}{30,-5}
\wlin{30,-5}{30,-30}
\flin{45,0}{30,20} \flin{30,20}{45,40}
\flin{30,-30}{45,-10} \flin{45,-50}{30,-30}
\Text(0,-25)[r]{\small 2} \Text(0,15)[r]{\small 1}
\Text(45,40)[l]{\small 3} \Text(45,0)[l]{\small 4}
\Text(45,-10)[l]{\small 5} \Text(45,-50)[l]{\small 6}
\Text(18,-65)[t]{\tt NONAB1}
\sof(0,-420)
\flin{0,40}{11,40} \lin{11,40}{33,40} \flin{33,40}{45,40}
\glin{13,40}{13,-5}{4}
\lin{13,-5}{13,-50}
\wlin{13,-5}{33,-5}
\flin{33,-5}{45,15} \flin{45,-25}{33,-5}
\flin{11,-50}{0,-50} \lin{33,-50}{11,-50} \flin{45,-50}{33,-50}
\Text(0,-50)[r]{\small 2} \Text(0,40)[r]{\small 1}
\Text(45,40)[l]{\small 3} \Text(45,15)[l]{\small 4}
\Text(45,-25)[l]{\small 5} \Text(45,-50)[l]{\small 6}
\Text(18,-65)[t]{\tt NONAB2}
\sof(80,-420)
\flin{0,40}{11,40} \lin{11,40}{33,40} \flin{33,40}{45,40}
\lin{13,40}{13,-5}
\lin{13,-5}{13,-50}
\glin{13,-5}{33,-5}{4}
\flin{33,-5}{45,15} \flin{45,-25}{33,-5}
\flin{11,-50}{0,-50} \lin{33,-50}{11,-50} \flin{45,-50}{33,-50}
\Text(0,-50)[r]{\small 2} \Text(0,40)[r]{\small 1}
\Text(45,40)[l]{\small 3} \Text(45,15)[l]{\small 4}
\Text(45,-25)[l]{\small 5} \Text(45,-50)[l]{\small 6}
\Text(18,-65)[t]{\tt NONAB3}
\sof(160,-420)
\flin{0,40}{11,40} \lin{11,40}{33,40} \flin{33,40}{45,40}
\lin{13,40}{13,-5}
\lin{13,-5}{13,-50}
\wlin{13,-5}{33,-5}
\flin{33,-5}{45,15} \flin{45,-25}{33,-5}
\flin{11,-50}{0,-50} \lin{33,-50}{11,-50} \flin{45,-50}{33,-50}
\Text(0,-50)[r]{\small 2} \Text(0,40)[r]{\small 1}
\Text(45,40)[l]{\small 3} \Text(45,15)[l]{\small 4}
\Text(45,-25)[l]{\small 5} \Text(45,-50)[l]{\small 6}
\Text(18,-65)[t]{\tt NONAB4}
\sof(240,-420)
\flin{0,15}{15,-5} \flin{15,-5}{0,-25}
\lin{15,-5}{30,-5}
\lin{30,20}{30,-5}
\lin{30,-5}{30,-30}
\flin{45,0}{30,20} \flin{30,20}{45,40}
\flin{30,-30}{45,-10} \flin{45,-50}{30,-30}
\Text(0,-25)[r]{\small 2} \Text(0,15)[r]{\small 1}
\Text(45,40)[l]{\small 3} \Text(45,0)[l]{\small 4}
\Text(45,-10)[l]{\small 5} \Text(45,-50)[l]{\small 6}
\Text(18,-65)[t]{\tt RAMB04}
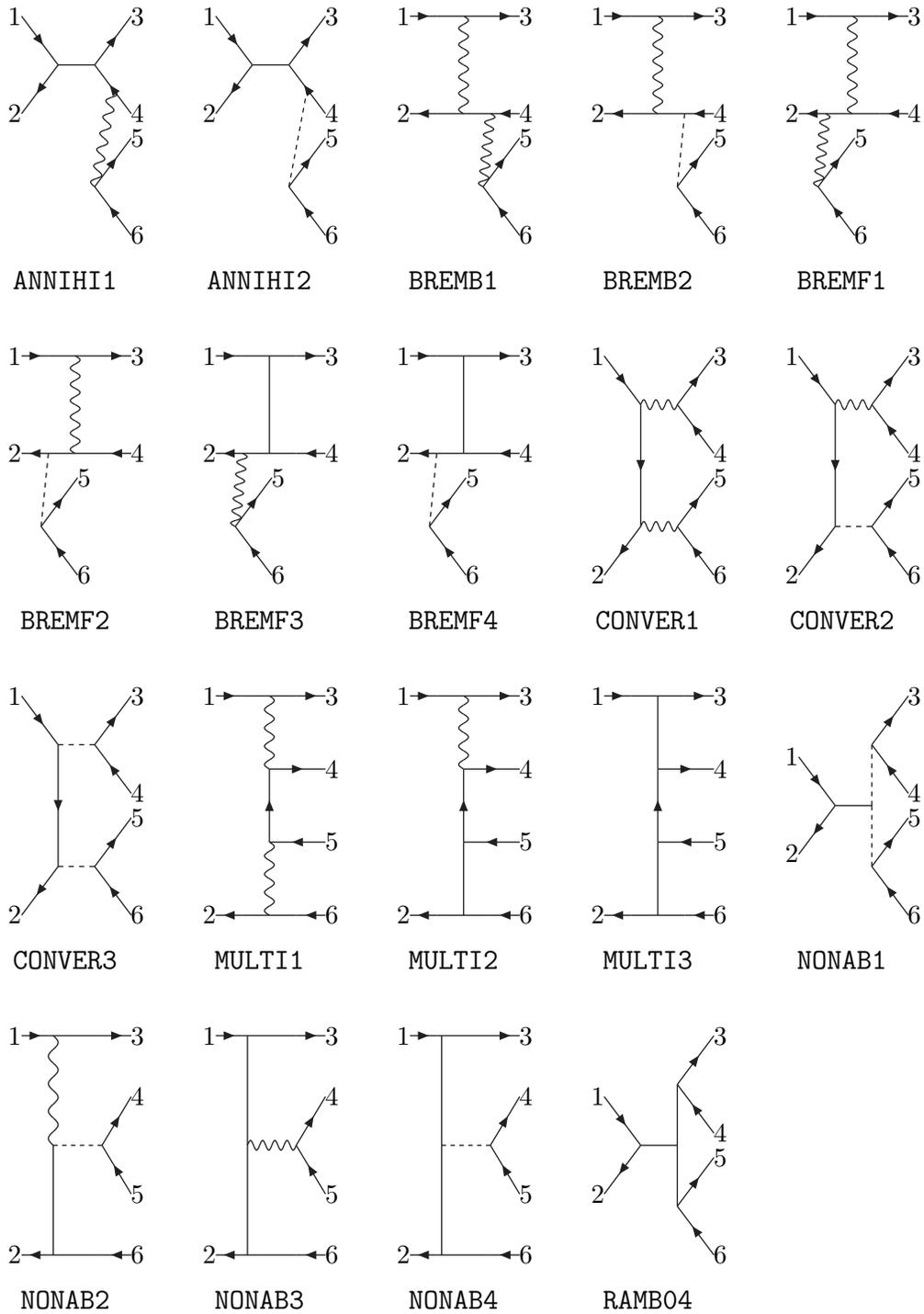
\Text(-80,-100)[]{Figure 2: kinematical diagrams in {\tt EXCALIBUR}.}
\end{picture}
\end{center}
\end{document}